\documentclass[twocolumn,showpacs,preprintnumbers,amsmath,amssymb,aps]{revtex4}
\usepackage{graphicx}
\usepackage{dcolumn}
\usepackage{bm}
\newcommand{\dxphi}{\frac{\partial \bar \phi}{\partial x}}
\newcommand{\pmean}{\bar p}
\newcommand{\pteq}{{\tilde p}^{eq}}
\newcommand{\vteq}{{\tilde v_x}^{eq}}
\newcommand{\cf}{Fig.~}

\begin{document}

\preprint{APS/123-QED}

\title{Control of transport barrier relaxations by resonant magnetic perturbations}

\author{M. Leconte$^{1}$, P. Beyer$^{1}$, X. Garbet$^2$ and S. Benkadda$^1$}
\affiliation{%
 $^1$ France-Japan Magnetic Fusion Laboratory LIA 336 CNRS / \\ UMR6633 CNRS - Universit\'{e} de Provence, Case 321, 13397 Marseille Cedex 20, France \\
 $^2$ CEA, IRFM, F-13108 Saint Paul Lez Durance, France.
  }%

\date{\today}

\begin{abstract}
Transport barrier relaxation oscillations, in presence of static Resonant Magnetic Perturbations (RMPs), are investigated using
3D global fluid turbulence simulations at the edge of a tokamak.
It is shown that RMPs have a stabilizing effect on these relaxation oscillations and that this effect is due mainly to a modification of the pressure profile linked to the presence of both residual magnetic island chains and a stochastic layer.
\end{abstract}

\maketitle


\indent
In magnetic fusion plasmas, turbulent transport reduces the energy confinement time, leading to a low confinement regime ($L$ mode). 30 years ago, the $H$ mode, a high confinement regime, linked to a sheared rotation of the plasma, was discovered. In this regime, the turbulence is locally reduced within a narrow layer, located a the plasma edge, known as a transport barrier. However, the transport barrier is not static but relaxes quasiperiodically, leading to violent bursts of heat and particles radially outward termed Edge Localized Modes (ELMs). The operationnal regime for next step fusion devices such as ITER will be Elmy H mode. ELMs are beneficial for power exhaust, but they represent however a threat for the plasma facing components and therefore need to be controlled.
Over the last decade, the possibility of controlling ELMs has become more and more plausible, as recent experiments were carried out on DIII-D using I-coils, on JET using error field correction coils and on the TEXTOR tokamak using the dynamic ergodic divertor \cite{Evans2004,Liang2007,Finken2007,Becoulet2008}. These experimental studies obtained a qualitative control over the ELMs by imposing static Resonant Magnetic perturbations (RMPs) at the plasma edge.
However, in order to get any quantitative result, work has to be done in the understanding of the interplay between transport barrier relaxations and RMPs.
Recently, elmy-like relaxation oscillations of transport barriers have been observed in global turbulence simulations of the tokamak edge \cite{Beyer2005, Beyer2007, Figarella2003, Fuhr2008}. The relaxation oscillations observed in these simulations have common characteristics with type-III ELMs. In this letter, we investigate numerically the effects of static Resonant Magnetic Perturbations (RMPs) on transport barrier relaxation oscillations.
The simulation domain is located in the edge region, around the $q=3$ resonant surface. We use the geometry of the TEXTOR tokamak, and plasma parameters close to those used in typical experiments on this machine \cite{Finken1999, Haberscheidt2006}.

The model used in this study describes electrostatic resistive ballooning turbulence involving the pressure $p$ and the electrostatic potential $\phi$.
The equations describing this model are the following \cite{Beyer2007}:
\begin{eqnarray}
\frac{\partial \nabla_\perp^2 \phi}{\partial t} + \{ \phi, \nabla_\perp^2 \phi \}
= & - \hat G p & -\nabla_\parallel^2 \phi + \nu \nabla_\perp^4 \phi
\label{rbm1}
\\
\frac{\partial p}{\partial t} + \{ \phi, p \}
= \delta_c \hat G \phi & + \chi_\parallel  \nabla_\parallel^2 p & + \chi_\perp \nabla_\perp^2 p + S
\label{rbm2}
\label{pressureeq}
\end{eqnarray}

The first equation corresponds to the vorticity equation, where $\nabla_\perp^2 \phi$ is the vorticity of the perpendicular (to the magnetic field) component of the $E\times B$ flow, and the parallel current and viscosity effects ($\nu$) are taken into account.
The second equation corresponds to energy conservation, where $\chi_\parallel$ and $\chi_\perp$  are collisional heat diffusivities parallel and perpendicular to the magnetic field, and $S=S(x)$ is an energy source modeling a constant heat flux density from the plasma core. Following the standard convention, $x$ represents the local radial coordinate, $y$ is the local poloidal coordinate and $z$ is the local toroidal coordinate, in a magnetic fusion device.
\\
The curvature operator $\hat G$ arises from the toroidal geometry of the tokamak, and $\delta_c=\frac{10}{3}\frac{L_p}{R_0} \ll 1$ is basically the ratio of the pressure gradient length $L_p$ to the tokamak major radius $R_0$. In the present model, time is normalized to the interchange time $\tau_{inter}=c_s^{-1}\sqrt{\frac{R_0 L_p}{2}}$, which also defines the perpendicular length scale through the ballooning length $\xi_{bal}=\sqrt{\frac{n_i^\textrm{ref} m_i~\eta_\parallel}{\tau_{inter}}}\frac{L_s}{B_0}$, where $c_s$ is the plasma sound speed, $B_0$ is the magnetic field strength, $n_i^\textrm{ref}$ is a reference ion density, $m_i$ is the ion mass, $\eta_\parallel$ is the parallel resistivity of the plasma and $L_s$ is a reference magnetic shear length. \\
In order to obtain a transport barrier, the poloidally and toroidally averaged component $\textbf{V}(x,t)=\dxphi~\textbf{e}_y$ of the $E\times B$ flow is driven by a forced poloidal sheared flow $\textbf{V}_F(x)=\Omega d \tanh(x/d)~ \textbf{e}_y$, chosen to be centered at $x=0$ (e.g. at the position of the main resonance of the RMPs), where $\Omega$ denotes the shear rate and  $d$ is the shear-layer width. Here, $\bar\phi(x,t)=\int_0^{2\pi}\int_0^{2\pi} \phi~dy dz$ is the poloidally and toroidally averaged electric potential.\\
Note that the parallel gradient is $\nabla_{\parallel}=\nabla_{\parallel0}+\{\psi_{RMP}~,~\cdot\}$, where $\nabla_{\parallel0}$ is the component due to the unperturbed magnetic field, $\{\psi_{RMP},\cdot \}= \frac{\partial\psi_{RMP} }{\partial x}\frac{\partial }{\partial y}-\frac{\partial \psi_{RMP}}{\partial y}\frac{\partial}{\partial x}$ denote the Poisson brackets. The magnetic flux due to the resonant magnetic perturbation is written as:
$
\psi_{RMP}(x,y,z)= \frac{I_D}{I_D^\textrm{ref}}\sum \psi_m(x) \cos\left( \frac{m}{r_0}y-\frac{n_0}{L_s}z \right)
$
where $ \psi_m (x)=\frac{\sin(m-m_0)}{m-m_0} \left(\frac{r_0}{r_c}\right)^m\exp\left({\frac{m}{r_0}x}\right)$ is the spectrum of the RMPs in slab geometry, $m_0$ is the central poloidal harmonic number, $r_c$ denotes the radial position of the RMP-producing coils, $I_D^\textrm{ref}$ is a reference value of the current in the RMP-producing coils, $n_0$ is the toroidal harmonic number and $r_0$ is a typical radius where the turbulence considered in this paper (resistive ballooning) develops. In the case studied here, we use $n_0=4$, with $q_{x=0}=q_0=3$, so that the central poloidal harmonic number is $m_0=12$.

Let the pressure be decomposed into a mean part $\bar p$ and harmonics $\tilde p=p-\bar p$, where $\bar p(x,t)=\int_0^{2\pi}\int_0^{2\pi}p~dy dz$ is the poloidally and toroidally averaged pressure. In a steady-state $\partial \bar p/\partial t\simeq 0$, the conservation of energy in the plasma (\ref{pressureeq}), where the conserved total energy flux is $Q_{tot}=\int S(x) dx=const.$, leads to the following equation:
\begin{equation}
 Q_{conv}
+ Q_{diff}
+ Q_{RMP}
= Q_{tot}
\label{energybalance}
\end{equation}

where $\displaystyle Q_{RMP}=\chi_{\parallel} \langle \frac{\partial\psi_{RMP}}{\partial y}\nabla_{\parallel}p \rangle$
represents the heat flux due to the magnetic flutter generated by the RMPs, and $Q_{conv}=\langle \tilde p \tilde v_x \rangle$ is the time-averaged convective heat flux, where $\tilde v_x$ denotes fluctuations of the radial velocity, and $Q_{diff}=-\chi_\perp d\bar p/dx$ is the diffusive heat flux. Here, $\langle \ldots \rangle$ denotes an average over the poloidal direction ($y$), the toroidal direction ($z$), and time $t$.

\begin{figure}
\begin{tabular}{cc}
\includegraphics[width=0.5\linewidth]{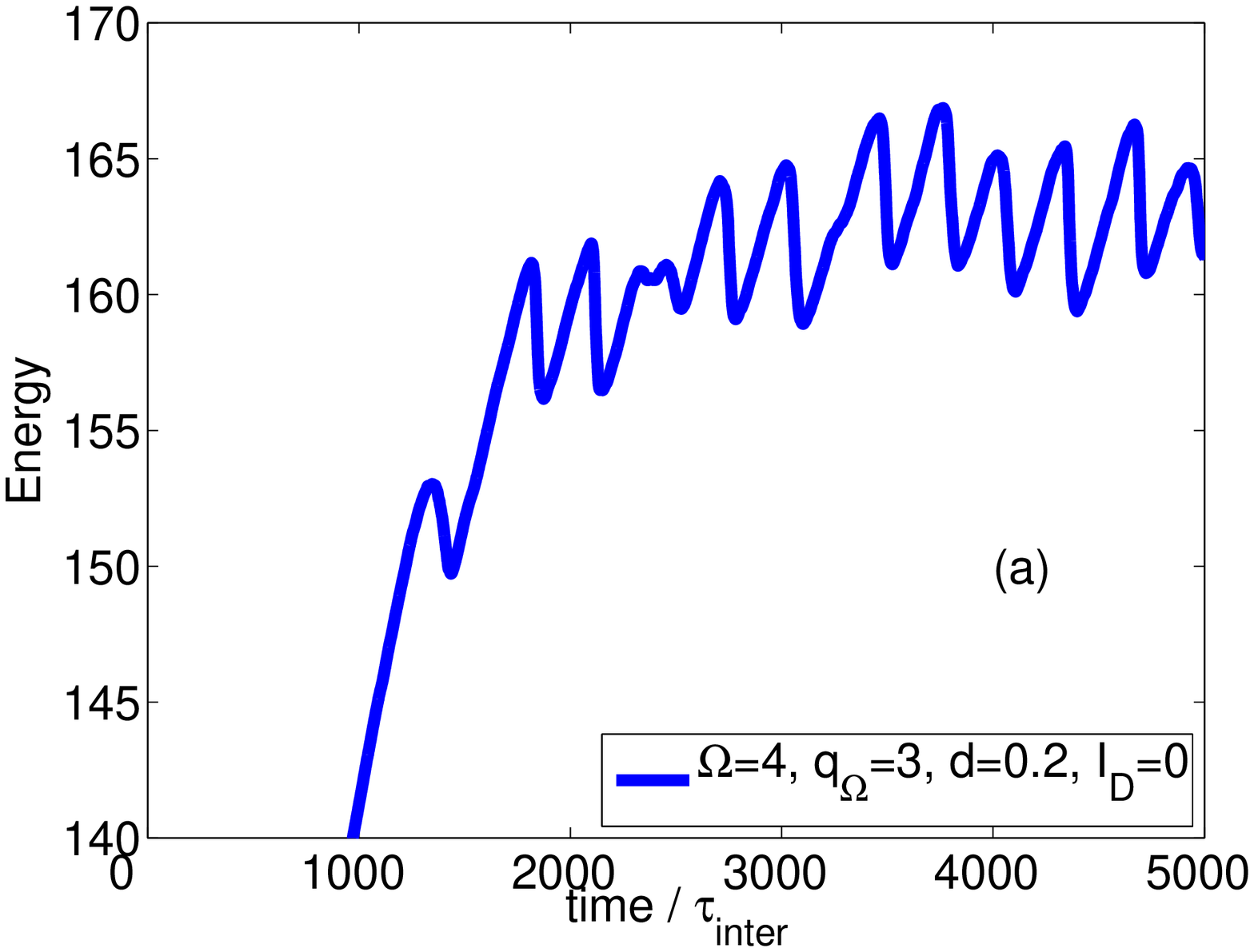}
\includegraphics[width=0.5\linewidth]{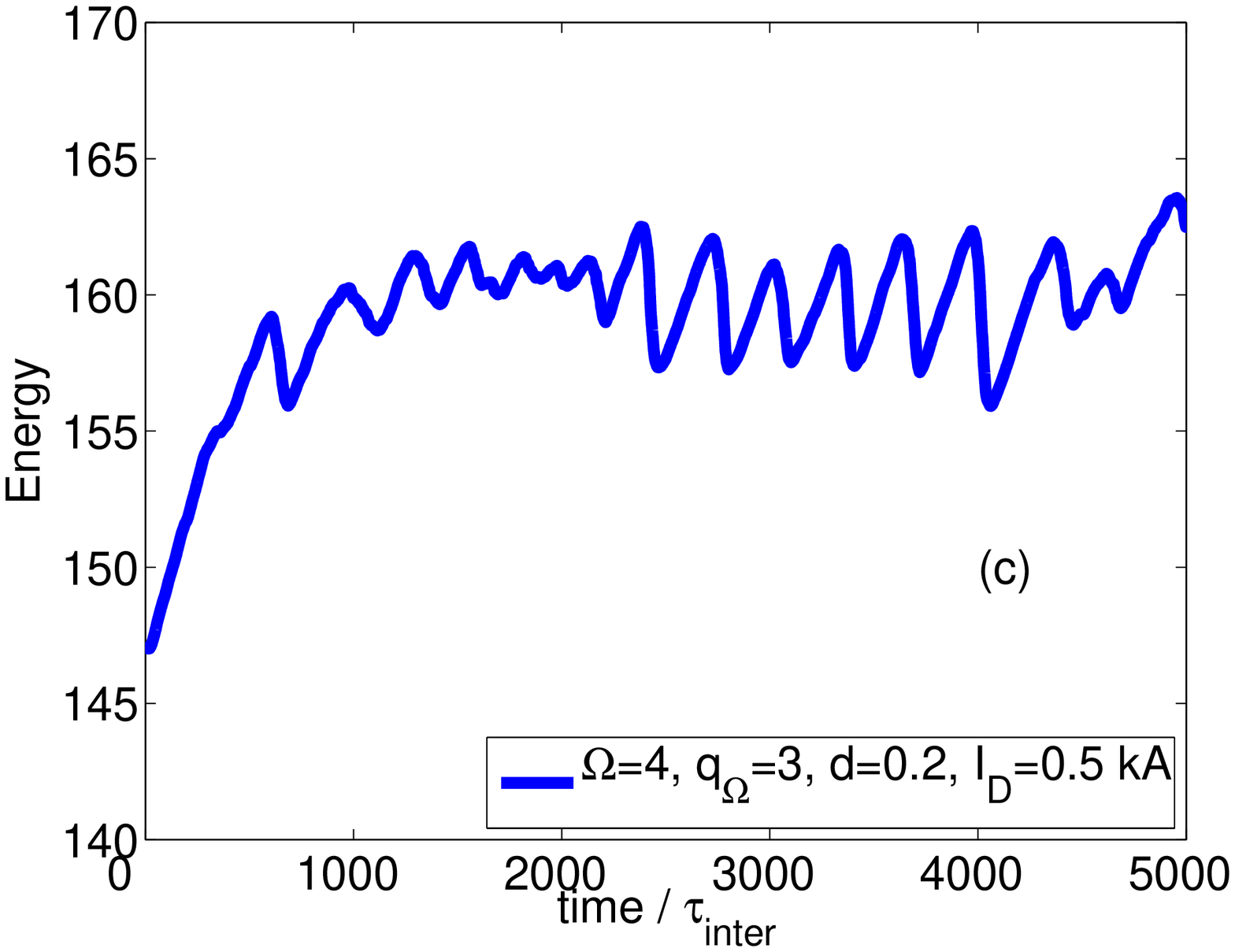} \\
\includegraphics[width=0.5\linewidth]{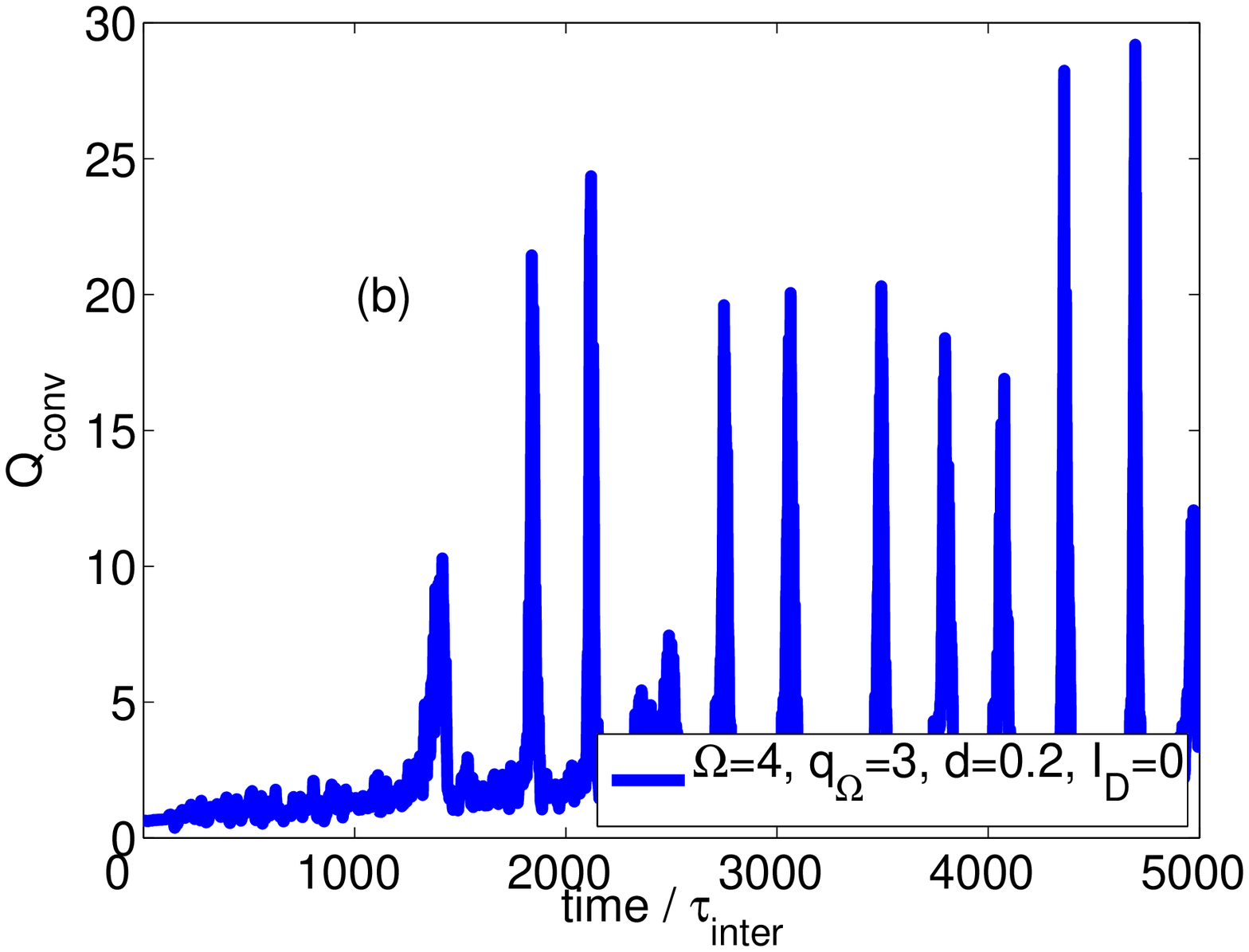}
 \includegraphics[width=0.5\linewidth]{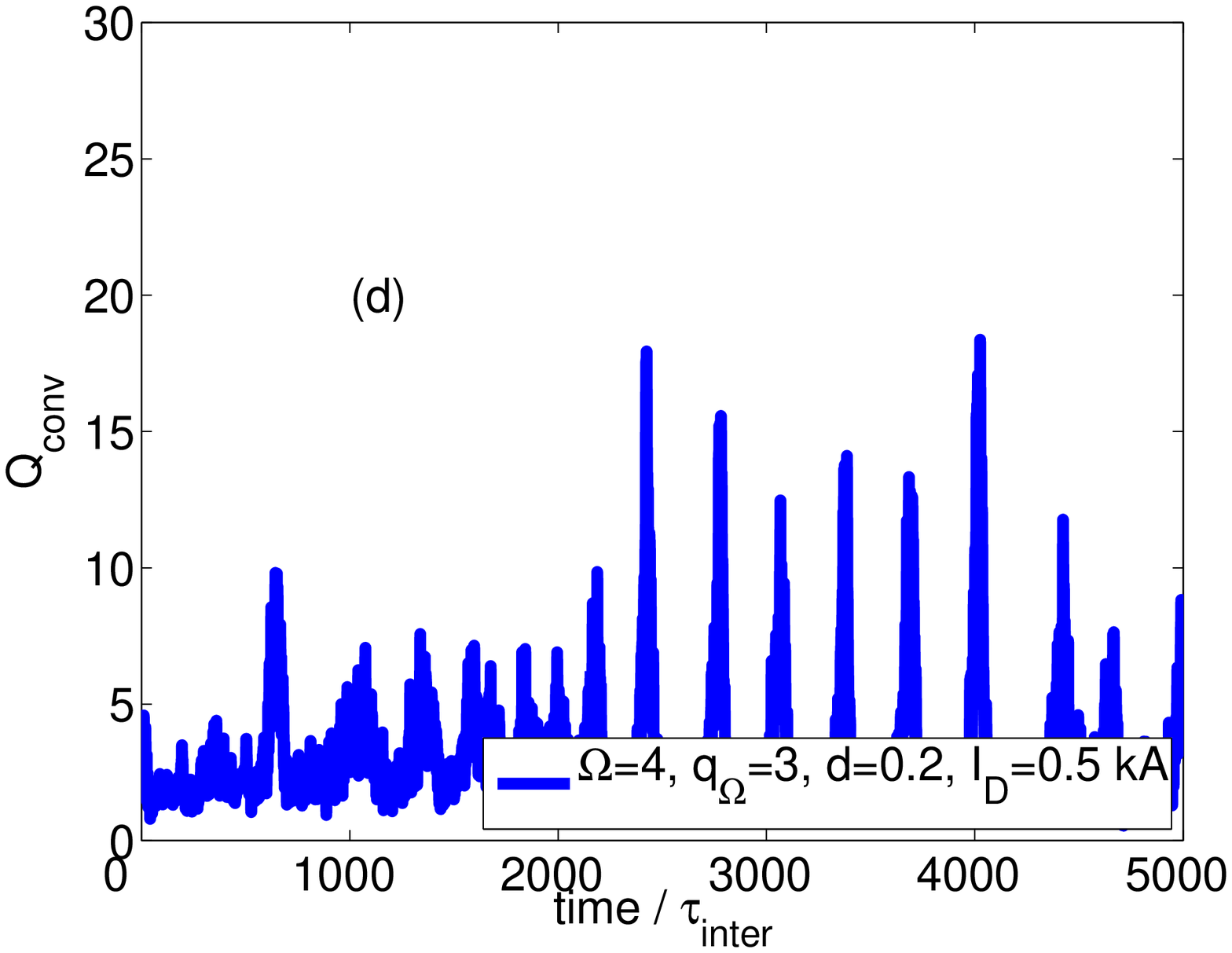}\\
\includegraphics[width=0.5\linewidth]{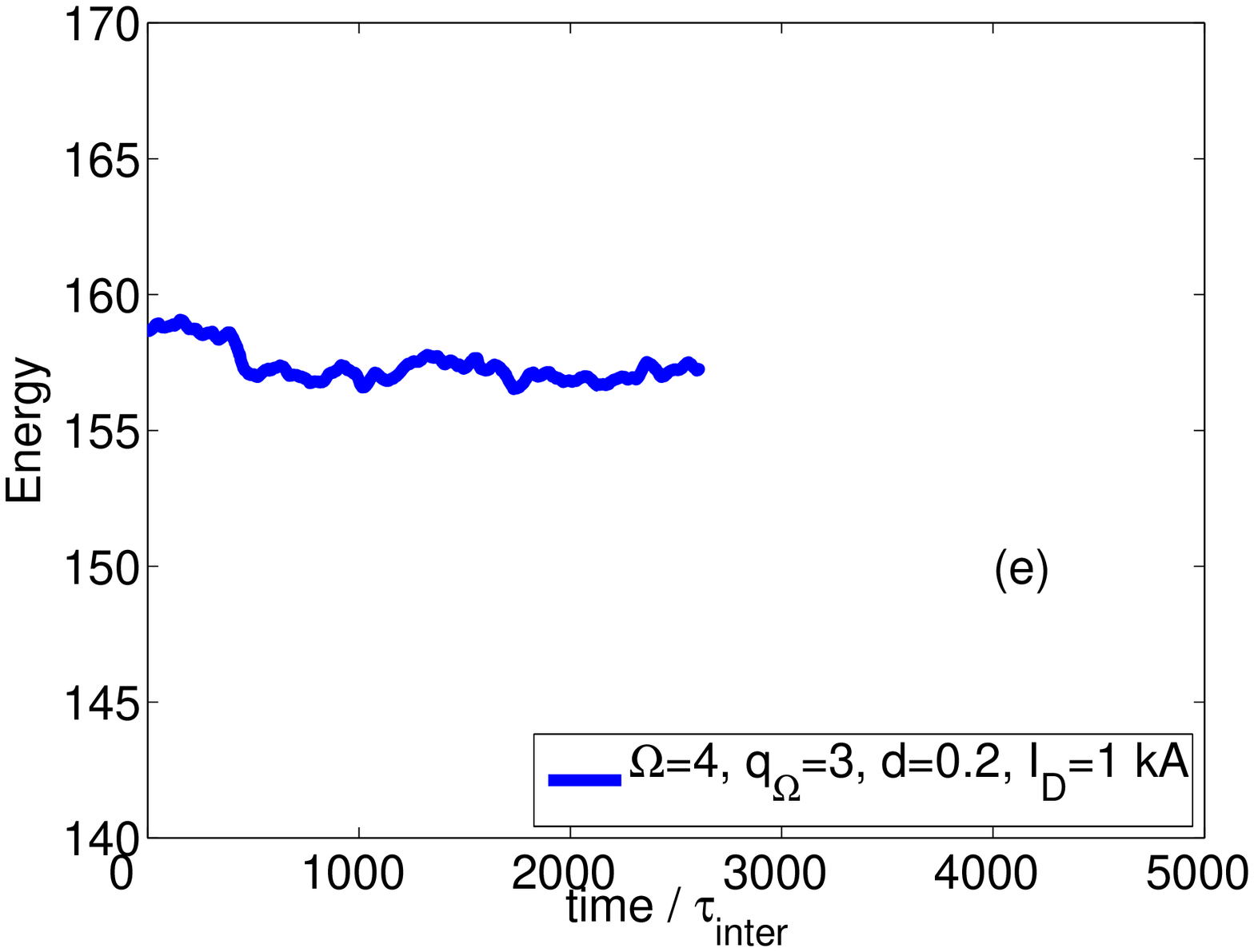} \includegraphics[width=0.5\linewidth]{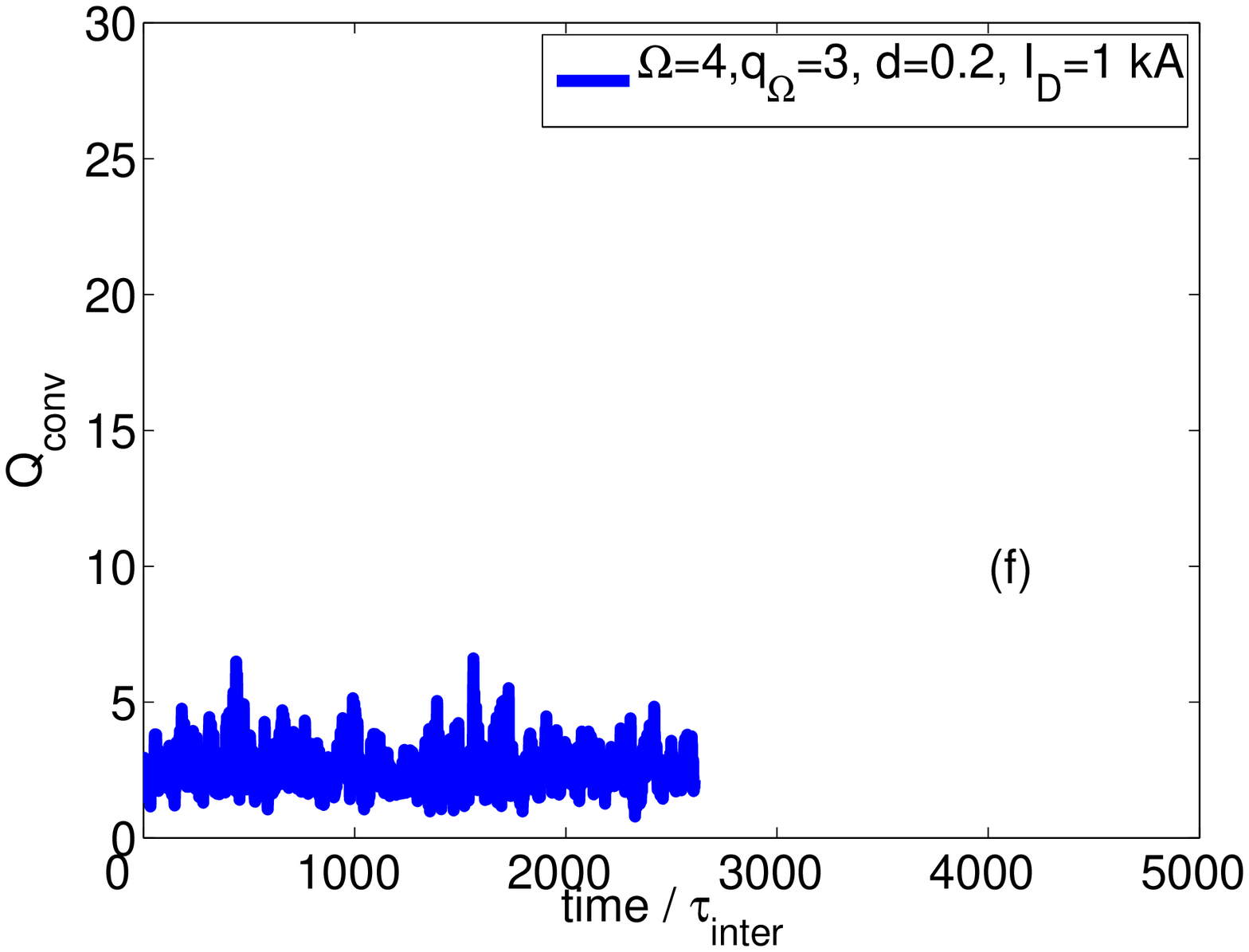}\\
\end{tabular}
	\caption{Effects of RMPs on the dynamics of barrier relaxations: time traces of the thermal energy (a,c,e) and the radial heat flux (b,d,f), in presence of an imposed mean shear flow with shear rate $\Omega=4$, for $(a,b)~$no RMP perturbation, $(c,d)~I_D=0.5~kA$ and $(e,f)~I_D=1~kA$.}
	\label{we4div0div1time}
\end{figure}

\indent Figure \ref{we4div0div1time} shows time series of the energy content $\int \bar p~dx$ and the convective flux $Q_{conv}$, in presence of an imposed mean sheared flow, for different values of the divertor current $I_D$.


In the reference case without RMPs [Fig\ref{we4div0div1time}a],
by imposing a forced sheared flow $\textbf{V}_F$ on the system, a steady-state is reached, where so-called relaxation oscillations are observed, corresponding to quasi-periodic relaxations of the pressure gradient. These relaxations are synchronous to the heat bursts observed on the heat-flux time series [\cf \ref{we4div0div1time}b] and therefore also correspond to relaxations of the transport barrier.


In the case with RMPs, the energy content shows that the relaxation oscillations are suppressed by the RMPs, and this suppression is more efficient for higher values of the perturbation current $I_D$ [\cf \ref{we4div0div1time}c,\ref{we4div0div1time}e]. This suppression of relaxations is also shown as a reduction of the amplitude of the heat bursts and an increase of their frequency [\cf \ref{we4div0div1time}d,\ref{we4div0div1time}f].

\begin{figure}
\begin{tabular}{c}
\includegraphics[width=0.5\linewidth]{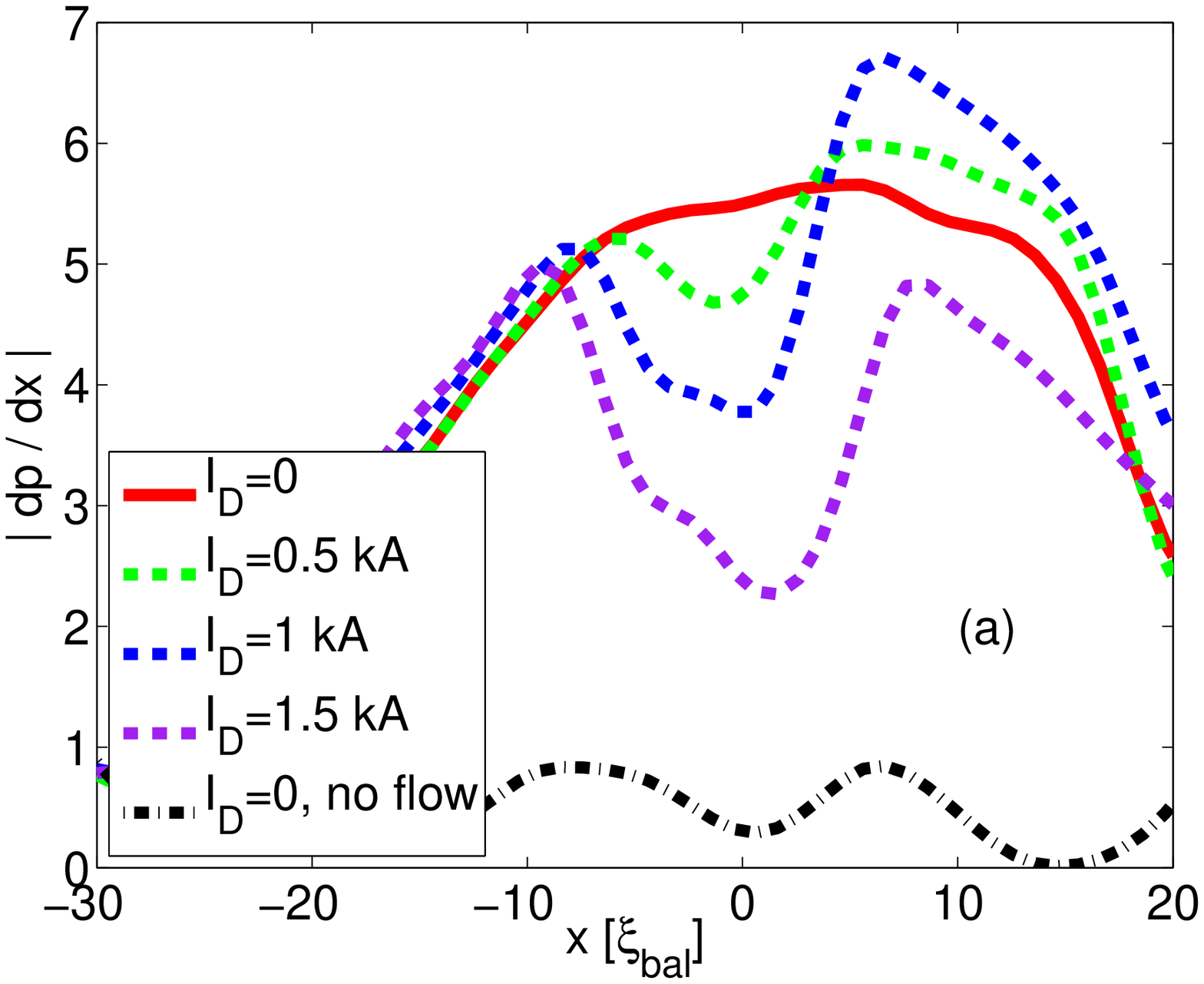}\\
\includegraphics[width=0.5\linewidth]{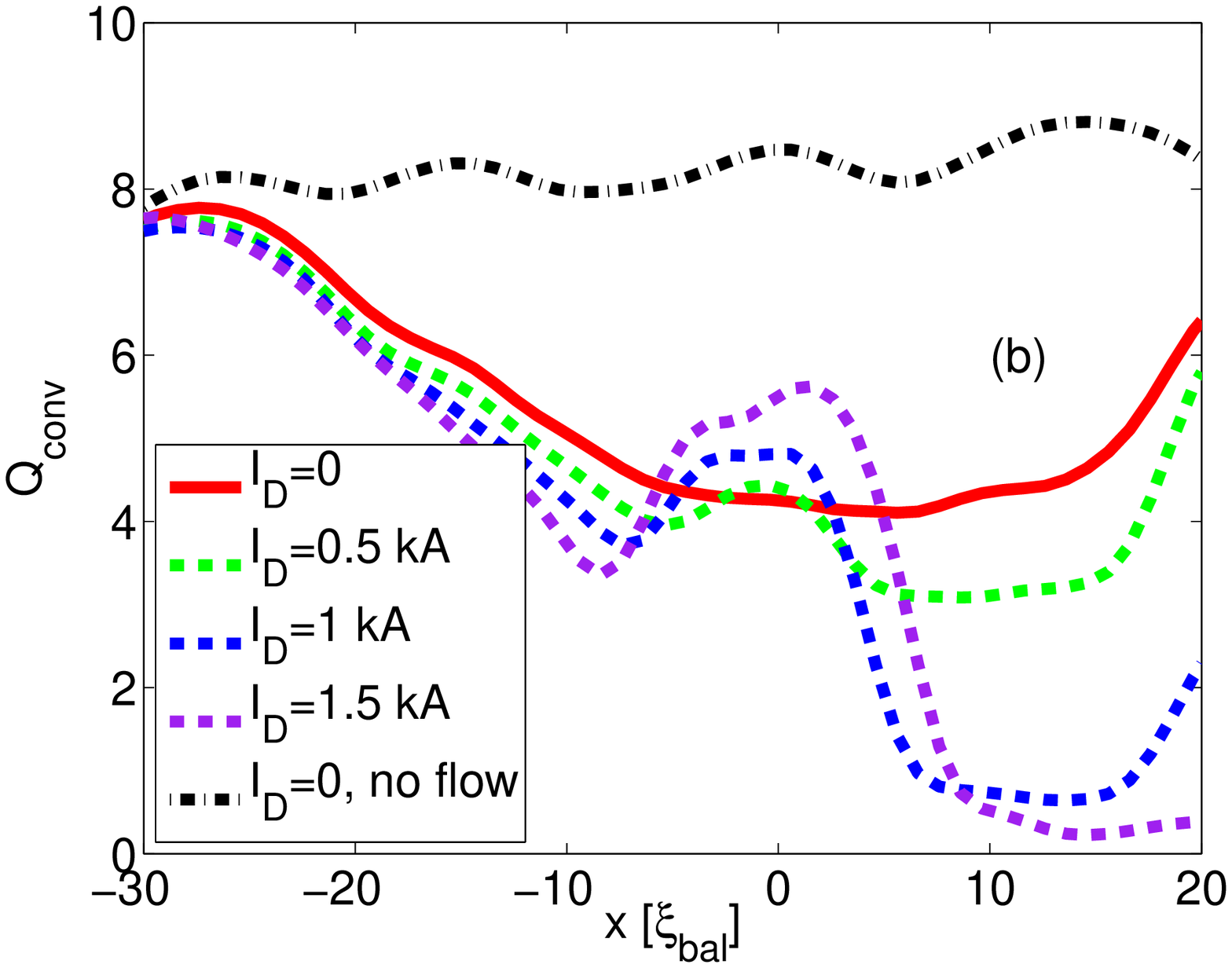}\\
\includegraphics[width=0.5\linewidth]{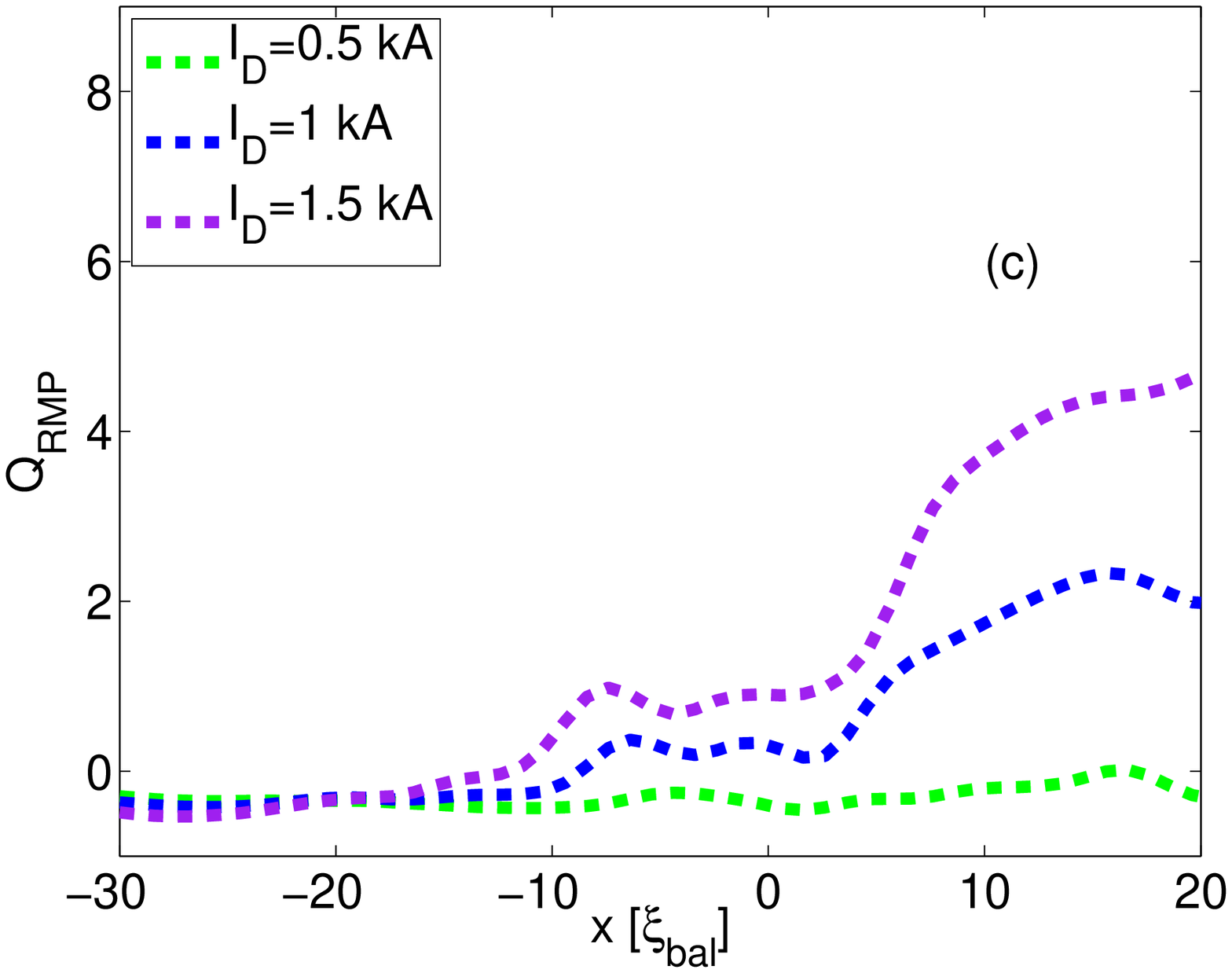}\\
\end{tabular}
	\caption{Effects of resonant magnetic perturbations on: (a) the pressure gradient and (b,c) radial heat fluxes, for different values of the perturbation current $I_D$. $Q_{conv}$ denotes the convective heat flux and $Q_{RMP}$ is the conductive heat flux induced by the resonant magnetic perturbations. The total heat flux is $Q_{tot}=10$. }
	\label{fluxes-space}
\end{figure}

Figure \ref{fluxes-space} shows the radial profiles of the pressure gradient $|\frac{d \pmean}{dx}|$, the convective flux $Q_{conv}$ and the RMP-induced flux $Q_{RMP}$, in presence of an imposed mean sheared flow, for different values of the divertor current $I_D$.
In the reference case without RMPs ($I_D=0$) but with a mean shear flow ($\Omega=4$), the convective flux $Q_{conv}$ [\cf \ref{fluxes-space}b] is reduced around the position $x=0$ compared with a case with no flow [\cf \ref{fluxes-space}b, dash-dotted line]. A high pressure gradient $|\frac{d \pmean}{dx}|$ around this position, corresponding to an ETB is created at the position of maximal flow-shear $x=0$ [\cf \ref{fluxes-space}a]. The appearance of this strong pressure gradient is linked to a reduction of the convective heat flux $Q_{conv}$ by the mean sheared flow
as seen from the conservation of energy (\ref{energybalance}) with $Q_{RMP}=0$:
\begin{equation}
Q_{conv}^0
+ Q_{diff}^0
=Q_{tot}
\label{energybalance0}
\end{equation}
where the superscript $0$ indicates the reference case without RMPs ($I_D=0$).\\



\indent In the case when there is a combination of a mean shear flow ($\Omega=4$) \emph{and} RMPs ($I_D \neq 0$), the ETB generated by the mean shear flow is eroded in the vicinity of the position $x\sim 0$, compared to the case without RMPs [\cf \ref{fluxes-space}a]. This erosion of the ETB only appears when both a mean shear flow and RMPs are present, and therefore can be explained by a synergetic effect.
p

\begin{figure}
\begin{tabular}{cc}
\includegraphics[width=0.5\linewidth]{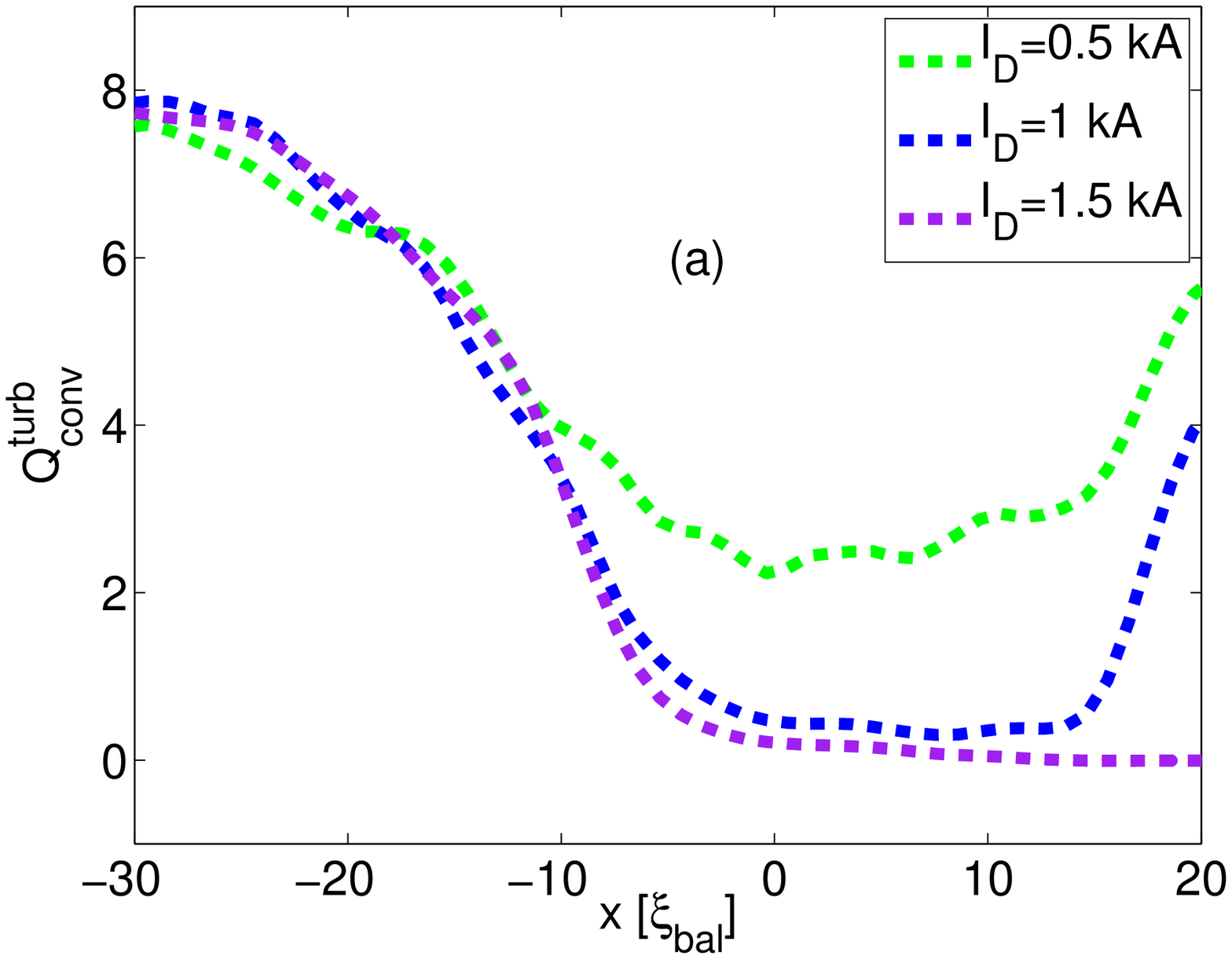} & \includegraphics[width=0.5\linewidth]{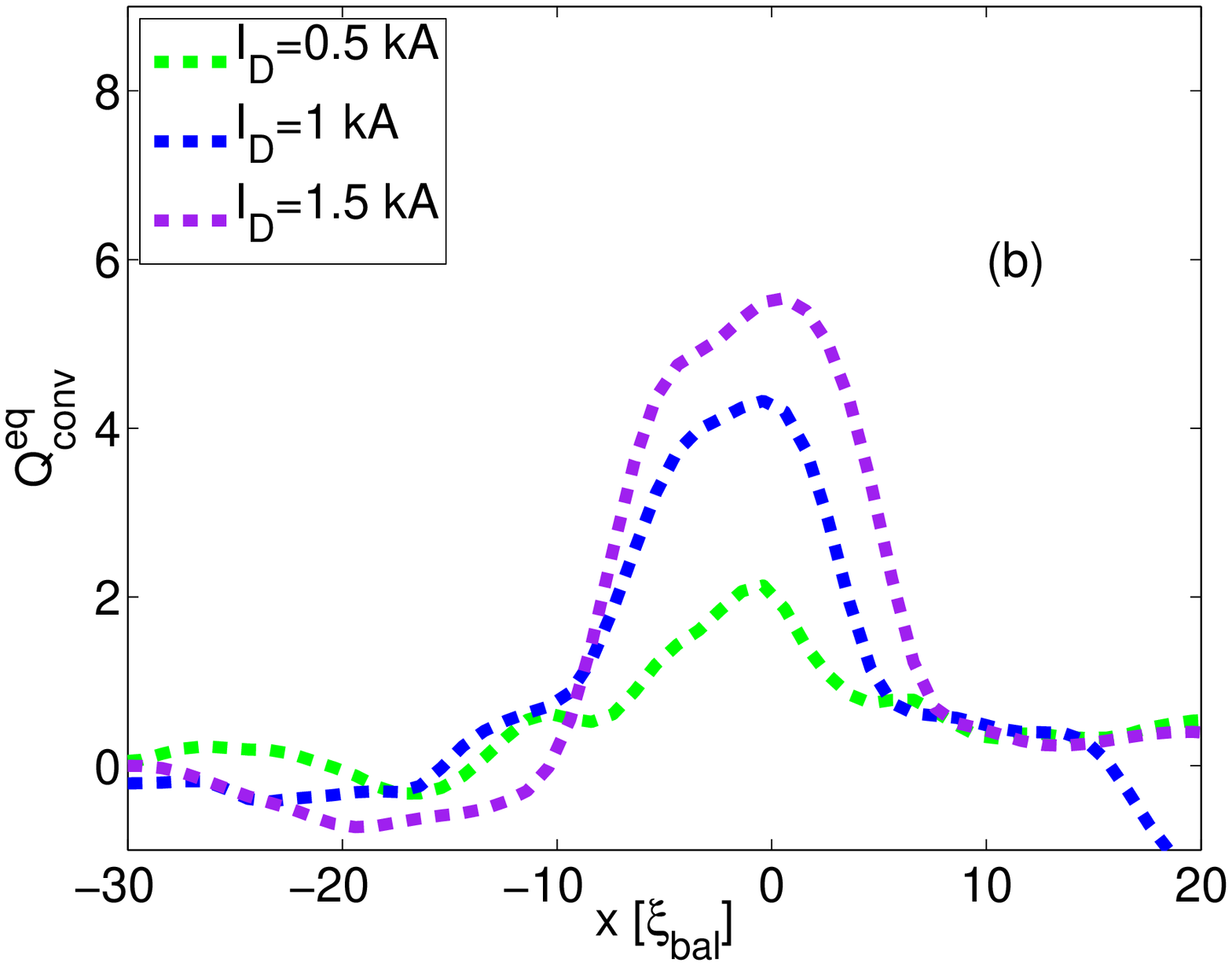} \\
\end{tabular}
\caption{Components of the radial convective flux: (a) turbulent convective flux $Q_{conv}^{turb}$, and (b) equilibrium convective flux $Q_{conv}^{eq}$, e.g. RMP-linked, for different values of the perturbation current $I_D$. The total heat flux is $Q_{tot}=10$.}
\label{conv2fluxes-space}
\end{figure}

We propose the following model based on the balance of heat fluxes, to explain the behaviour of the convective flux and pressure gradient in the presence of RMPs and a shearflow-induced transport barrier.
Taking into account RMPs, it is convenient to decompose the pressure harmonics into an equilibrium and a turbulent part: $\tilde p=\tilde p^{eq}(x,y,z)+\tilde p^{turb}(x,y,z,t)$, and similarly for the radial velocity harmonics $\tilde v_x$. The energy balance (\ref{energybalance}) can then be written:
\begin{equation}
Q_{conv}^{eq} + Q_{conv}^{turb}
+ Q_{diff}
+ Q_{RMP}
=Q_{tot}
\label{energybalance1}
\end{equation}
where $Q_{conv}^{eq}=\langle \pteq \vteq \rangle$ is the equilibrium convective flux and $Q_{conv}^{turb}=\langle \tilde p^{turb} \tilde v_x^{turb} \rangle$ is the time averaged convective turbulent flux.

\begin{figure}
\begin{tabular}{cc}
\includegraphics[width=\linewidth]{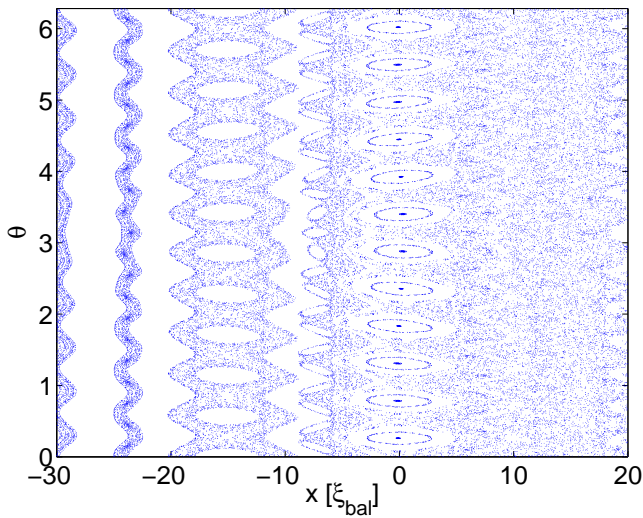}
\\
\end{tabular}
	\caption{Effects of RMPs on the topology of magnetic field lines: Poincare map for a perturbation current $I_D=0.5~kA$. There is clear evidence of three residual magnetic island chains, and of a stochastic layer in between the central island chain and the one on the right.}
	\label{poincare}
\end{figure}

Because of the presence of localized residual island chains (as seen on Figure \ref{poincare}) of normalized width $w\sim \sqrt{I_D/I_D^\textrm{ref}}$, depending on the region considered, there are three different limits:

i) Close to the resonant surface $r=r_0$, e.g. $|x| \ll \frac{w}{2} \ll d$, the turbulent convective flux is reduced by the sheared flow and additionnally reduced by RMPs due to stochasticity of magnetic field lines [\cf \ref{poincare}] so that $Q_{conv}^{turb} /Q_{tot} \ll 1$ [\cf \ref{conv2fluxes-space}a]. Moreover, an equilibrium convective flux $Q_{conv}^{eq}$ indirectly linked to RMPs appears [\cf \ref{conv2fluxes-space}b]. This equilibrium convective flux is a consequence of the presence of an RMP-induced residual magnetic island chain at the radius $r=r_0$ (in TEXTOR, $r_0=0.45~[m]$) corresponding to the position $x=0$ [\cf \ref{poincare}]. Moreover, the RMP induced flux is low in the vicinity of $x=0$: $Q_{RMP}/Q_{tot} \ll 1$ [\cf \ref{fluxes-space}c]. Therefore, the energy balance (\ref{energybalance1}) simplifies to:
\begin{equation}
Q_{conv}^{eq} 
+ Q_{diff}
  \sim Q_{tot}
\end{equation}
so that, in the region $x \sim 0$, the appearance of the equilibrium convective flux $Q_{conv}^{eq}$ must be balanced by a decrease of the pressure gradient, seen on Figure \ref{fluxes-space}a, similar to a flattening of the pressure profile on the island chain, that occurs in the study of plasma macroinstabilities such as tearing modes \cite{Fitzpatrick95}.\\
ii) For $r<r_0$, far from the resonant surface but inside the barrier region, e.g. $-d \ll x \ll -\frac{w}{2}$, the RMP-linked equilibrium convective flux is small $Q_{conv}^{eq}/Q_{tot} \ll 1$, since there is no residual island chain in this region, so it does not play any role. The turbulent convective flux $Q_{conv}^{turb}$ is reduced by the mean-shear flow. Moreover, the RMP-induced flux $Q_{RMP}$ is small in this region $Q_{RMP} /Q_{tot} \ll 1$. Therefore, the energy balance (\ref{energybalance1}) reduces to:
\begin{equation}
Q_{conv}^{turb}
+ Q_{diff}
\sim Q_{tot}
\label{energybalance2}
\end{equation}

Additionnally, in this region, the turbulent convective flux does not depend on the perturbation current $I_D$ [\cf \ref{conv2fluxes-space}a] because the magnetic field lines are not stochastic [\cf \ref{poincare}], so that for $-d \ll x \ll -\frac{w}{2}$ the turbulent convective flux is only weakly perturbed: $Q_{conv}^{turb} \sim Q_{conv}^0$. Therefore, equations (\ref{energybalance0}) and (\ref{energybalance2}) imply:
$
Q_{diff} \sim Q_{diff}^0
$.
Thus the pressure gradient profile is only weakly modified by the RMPs [\cf \ref{fluxes-space}a]. \\
iii) For $r>r_0$, far from the resonant surface but inside the barrier region, e.g. $\frac{w}{2} \ll x \ll d$, the equilibrium convective flux is small ($ Q_{conv}^{eq}/Q_{tot} \ll 1$), since there are no residual island chains. The energy balance (\ref{energybalance}) thus reduces to:
\begin{equation}
Q_{conv}^{turb}
+ Q_{diff}
+ Q_{RMP}
\sim Q_{tot}
\end{equation}
An increase of the perturbation current $I_D$ causes an increase of the RMP-induced heat flux $Q_{RMP}$, but it also induces a decrease of the turbulent convective heat flux $Q_{conv}^{turb}$, linked to stochasticity of magnetic field lines [\cf \ref{poincare}].
In this region, there is therefore a competition between the $Q_{RMP}$ and $Q_{conv}^{turb}$ heat fluxes, which may explain the fact that, in the region $\frac{w}{2} \ll x \ll d$, the pressure gradient increases for small perturbation currents and decreases for higher perturbation currents [\cf \ref{fluxes-space}a]. \\

\begin{figure}
\begin{tabular}{cc}
\includegraphics[width=0.5\linewidth]{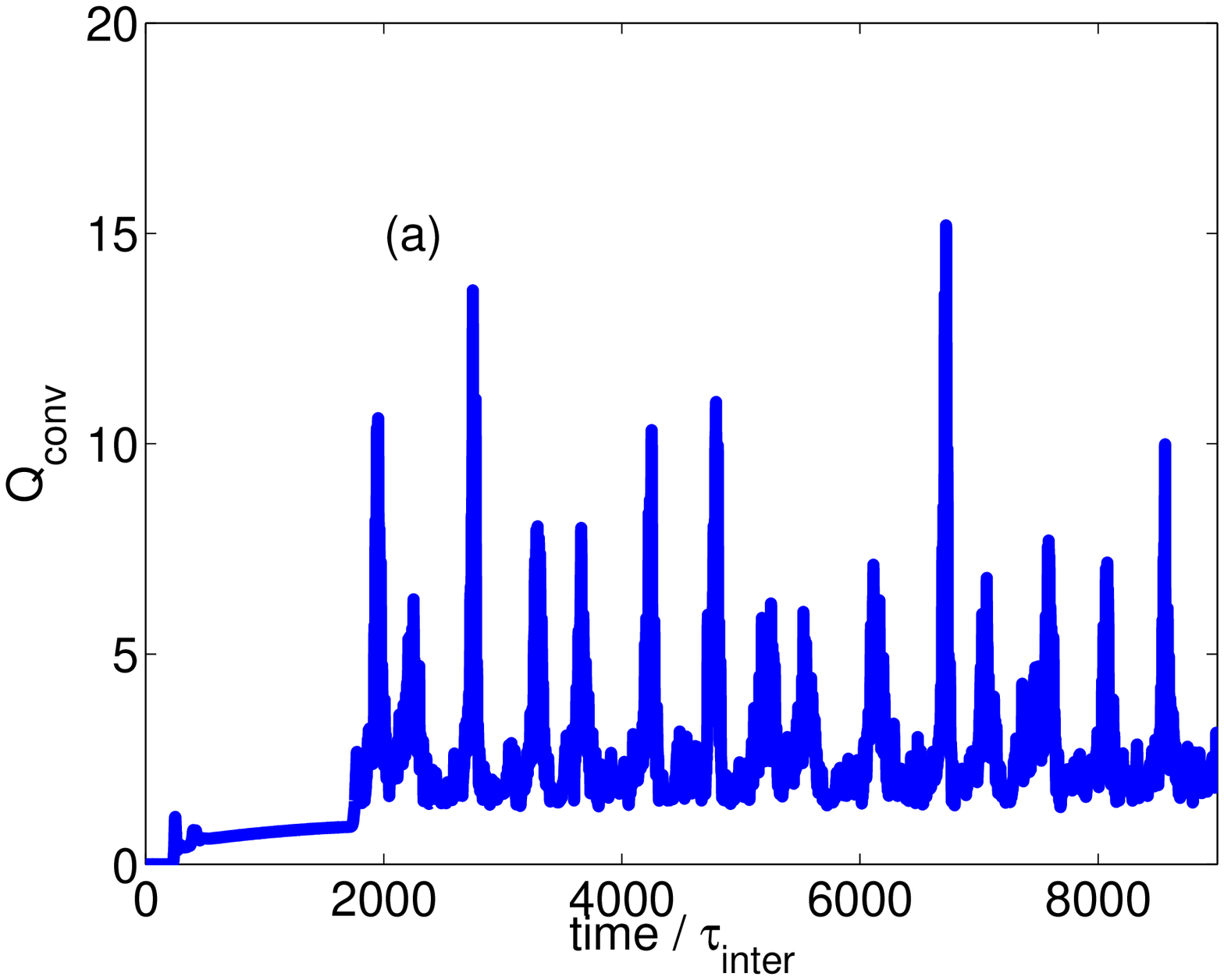}
 & \includegraphics[width=0.5\linewidth]{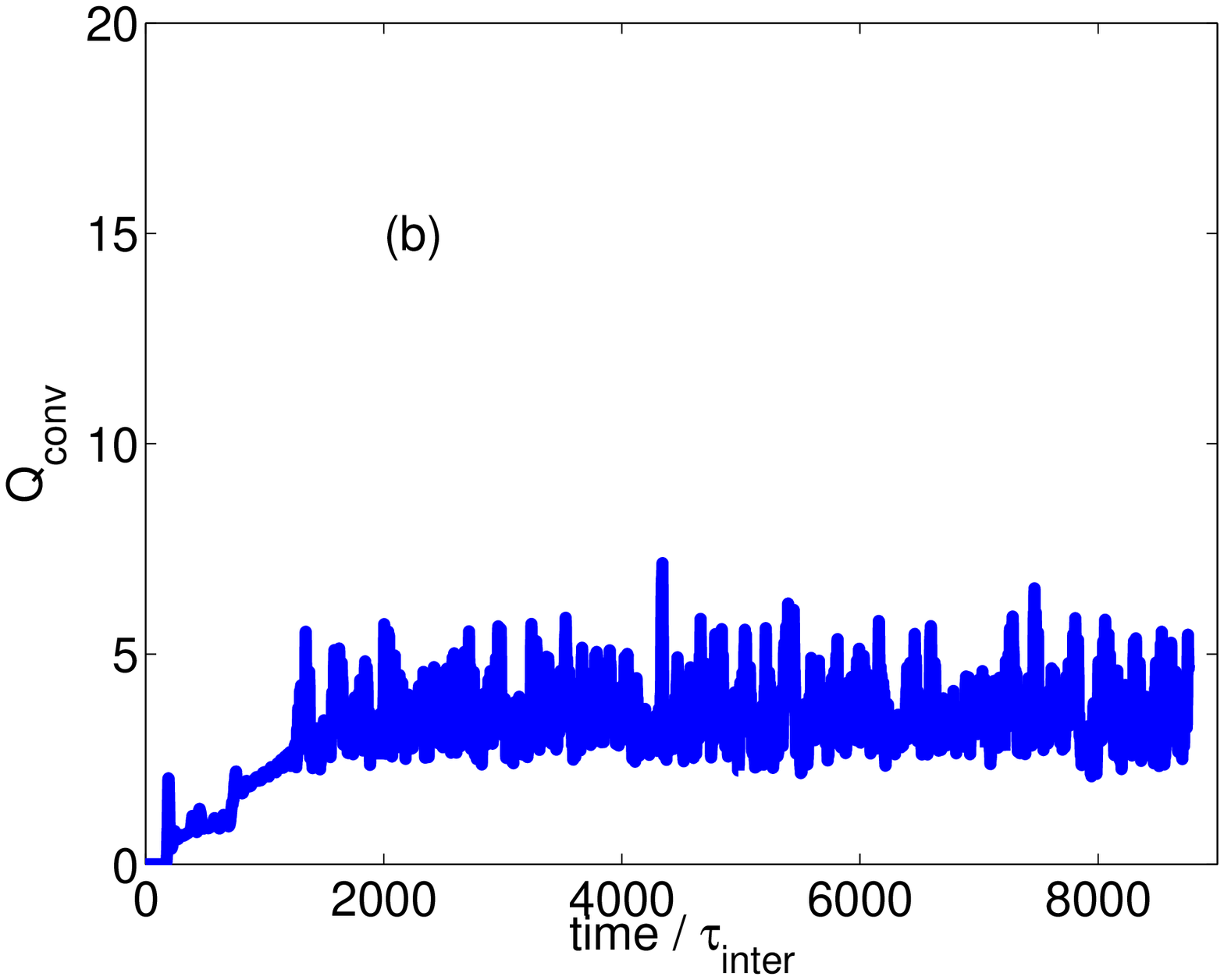}\\
\end{tabular}
	\caption{Effects of width $d$ of the shear-layer on the dynamics of barrier relaxations: time traces of the radial turbulent flux, in presence of an imposed mean sheared flow with shear rate $\Omega=2$, for $a)~d=20\%$ and $b)~d=10\%$.}
	\label{widthtime}
\end{figure}

From the analysis presented here, the stabilization of relaxation oscillations appears to be linked to an erosion of the transport barrier, i.e. a reduction of its width.

To confort this hypothesis, we performed simulations without RMPs, but with a narrower transport barrier (produced by a sheared flow with a thinner shear-layer $d$). Figure \ref{widthtime} shows time series of the convective flux, in the case without RMPs, for two different values of the shear-layer width $d$.
As seen from the comparison of \cf \ref{we4div0div1time} and \cf \ref{widthtime}, there is a striking similarity between the effects of a decrease of the shear-layer width $d$ \emph{and} the effects of RMPs on the dynamics of Edge Transport Barrier (ETB) relaxations. Our simulations therefore suggests that the main effect of RMPs in presence of an ETB is to modify the geometrical properties of the ETB, e.g its width and position, yielding a reduction of the amplitude and frequency of the elmy-like relaxations, and therefore leading to grassy elmy-like relaxations.\\
To summarize, we investigated the effects of resonant magnetic perturbations (RMPs) on transport barrier relaxations.
In conclusion, it is shown that RMPs have a stabilizing effect on these relaxations. This effect is linked to a modification of the pressure gradient equilibrium profile due mainly to a modification of the magnetic topology, e.g. the formation of magnetic island chains, and to RMP-induced stochastic transport. An erosion of the pressure gradient profile is observed at the surface of principal resonance (also chosen to be the central position of the sheared flow). This erosion is shown to be linked to the presence of residual magnetic island chains inducing a stationnary convective transport of heat (and particles) in the radial direction. Far from the principal resonance surface but inside the shear-layer, the pressure gradient modifications are only linked to the presence (or not) of stochastic resonance overlap. 


\end{document}